\documentclass[traditabstract]{aa}
\usepackage[dvips]{graphicx}
\usepackage{txfonts}
\usepackage{natbib}
\bibpunct{(}{)}{;}{a}{}{,}

\begin{document}

\title{The first WASP public data release}

\author{
  O. W. Butters\inst{1}
\and  
  R. G. West\inst{1}
\and
  D. R. Anderson\inst{2}
\and
  A. Collier Cameron\inst{3}
\and
  W. I. Clarkson\inst{4}
\and
  B. Enoch\inst{3}
\and
  C. A. Haswell\inst{5}
\and
  C. Hellier\inst{2}
\and
  K. Horne\inst{3}
\and
  Y. Joshi\inst{6}
\and
  S. R. Kane\inst{7}
\and
  T. A. Lister\inst{8}
\and
  P. F. L. Maxted\inst{2}
\and
  N. Parley\inst{3}
\and
  D. Pollacco\inst{6}
\and
  B. Smalley\inst{2}
\and
  R. A. Street\inst{8}
\and
  I. Todd\inst{6}
\and
  P. J. Wheatley\inst{9}
\and
  D. M. Wilson\inst{2}
}

\institute{
  Department of Physics and Astronomy, University of Leicester,
  Leicester, LE1 7RH, U.K.\\
  \email{oliver.butters@star.le.ac.uk}
\and
  Astrophysics Group, School of Physical and Geographical Sciences,
  Lennard-Jones Laboratories, Keele University, Staffordshire, ST5
  5BG, U.K.
\and
  School of Physics and Astronomy, University of St Andrews, North
  Haugh, St Andrews, Fife, KY16 9SS, U.K.
\and
  STScI, 3700 San Martin Drive, Baltimore, MD 21218, U.S.A.
\and
  Department of Physics and Astronomy, The Open University, Milton
  Keynes, MK7 6AA, U.K.
\and
  Astrophysics Research Centre, Main Physics Building, School of
  Mathematics \& Physics, Queen's University, University Road, Belfast,
  BT7 1NN, U.K.
\and
  NASA Exoplanet Science Institue, Caltech, Pasadena, CA 91125, U.S.A.
\and
  Las Cumbres Observatory, 6740 Cortona Drive Suite 102, Goleta, CA
  93117, U.S.A.
\and
  Department of Physics, University of Warwick, Coventry, CV4 7AL, U.K.
}

\authorrunning{Butters et al.}

\date{Accepted 2007 ???; Received  2007 ???}

\abstract{The WASP (Wide Angle Search for Planets) project is an
  exoplanet transit survey that has been automatically taking wide
  field images since 2004. Two instruments, one in La Palma and the
  other in South Africa, continually monitor the night sky, building up
  light curves of millions of unique objects. These light curves are
  used to search for the characteristics of exoplanetary transits.
  This first public data release (DR1) of the WASP archive makes
  available all the light curve data and images from 2004 up to 2008 in
  both the Northern and Southern hemispheres. A web interface
  (\texttt{www.wasp.le.ac.uk/public/}) to the data allows easy
  access over the Internet. The data set contains 3\,631\,972 raw images and
  17\,970\,937 light curves. In total the light curves have
  119\,930\,299\,362 data points available between them.}


\maketitle

\section{Introduction}

There have been over 450 exoplanets discovered to date, over 60 of
these transit their host star\footnote{\texttt{www.exoplanet.eu}}. The
WASP (Wide Angle Search for Planets) project has played a
pivotal role in the field by finding almost half of these transiting
systems. This has been achieved by the construction of two robotic
observatories (one in the Northern hemisphere and the other in the
Southern hemisphere) that constantly monitor the night sky. Images are
taken simultaneously from eight wide-angle cameras in each case
throughout the night.

The first science frames were taken in 2004 and since then well over 200 billion data
points have been taken. This forms the basis of the main data product
offered here; that of light curves of individual stars. These light
curves have been extensively searched for the characteristic signs of
transiting exoplanets, and to date over 30 have been found.

The format of the data lends itself to non-exoplanet research also,
for example variable star studies
\citep{norton07} and single star studies \citep{cameron09}. This
dataset will therefore be a valuable resource to the wider community
of time domain astronomers.

This first public data release (DR1) represents over half of the
current total WASP data holdings, and will be followed with further
releases covering more of the sky and a longer temporal baseline.

The remainder of this paper is organised as
follows; Section~\ref{SW_observatories} describes the WASP
observatories. Section~\ref{data_summary} gives an overview of the
data available. Section~\ref{data_products} outlines the data products
that are available. Section~\ref{interface} gives an
overview of the web interface to the data.

\section{The WASP observatories}
\label{SW_observatories}

\begin{figure}
  \includegraphics[width=\hsize]{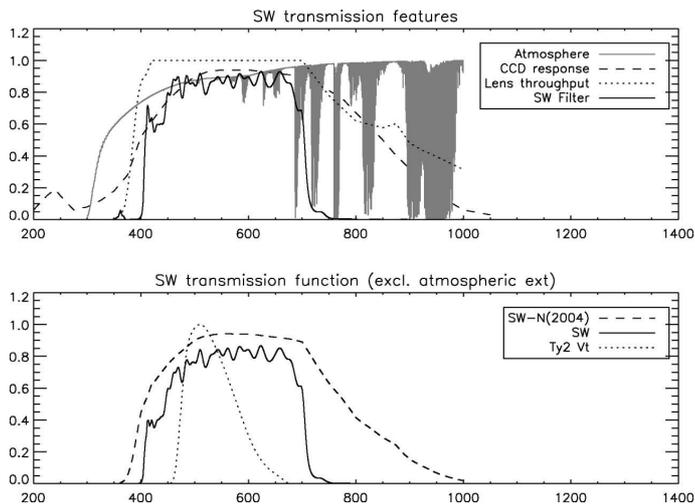}
  \caption{Passband of the WASP filter (top) plotted alongside
    the atmospheric transmission, CCD response, and lens
    transmission. The bottom panel shows the original unfiltered system
  alongside the current filter. (Taken from \citet{pollacco06}).}
  \label{passband}
\end{figure}

The WASP observatories consist of two identical robotic
telescopes, one located at the Observatorio del Roque de los Muchachos
on La Palma, the other at the South African Astronomical Observatory
\citep{pollacco06,wilson08}. Each telescope has eight lenses (Canon
200mm f/1.8) feeding a 2048$\times$2048 thinned CCD with a pixel size
of 13.5$\mu$m. This gives a field of view of 7.8$\times$7.8 degrees
(61 sq. degrees) per camera and an angular scale of
13.7$^{\prime \prime}$~pixel$^{-1}$. The cameras are fixed relative to
one another on an equatorial mount. In 2004 the light was unfiltered
- the spectral transmission essentially being defined by the optics,
detectors and the atmosphere. From 2006 onwards a broadband filter was
installed with a passband from 400 to 700~nm (see
Fig.~\ref{passband}).

Each night calibration frames are taken at the beginning and end of the
night, then the exoplanet survey runs throughout most of the rest of
the night. The observing strategy is optimised to cycle through 6-8
fields at a similar declination and spaced apart by approximately one
hour in right ascension each night. This strategy is altered somewhat
to avoid crowded fields close to the Galactic plane. Two exposures are
taken of each field (each 30 seconds) then the telescope slews to
the next field (taking approximately 30 seconds), so each field is
sampled every 9-12 minutes. Twice per night
the exoplanet survey is interrupted to perform a full sky survey, a
process that takes approximately 40 minutes.

The raw images are processed by Queen's and Keele Universities (North
and South data respectively) using a custom built pipeline (see
\citet{kane04} and \citet{pollacco06}). The final
result of this analysis is a photometric measurement of each object in
each image. This data is then ingested into the main archive hosted at
the University of Leicester, where it is subject to high-level
analysis, such as de-trending, and searched for exoplanet signatures.

The data reduction pipeline is catalogue driven and backed by the
USNO-B1.0 catalogue \citep{monet03}. Each object extracted from the
images has its position matched against the catalogue, when a match is
found it is ingested into the main archive, when it is not it is
ingested into a separate part of the archive. The sources that were
successfully matched against the USNO-B1.0 catalogue are the ones
available here, so transient sources are therefore unlikely to be seen
in this data release. More details about the pipeline can be found in
\citet{pollacco06}.

\section{Data summary}
\label{data_summary}

This release has 3\,631\,972 raw science images available for download. From
these images 17\,970\,937 unique objects have been catalogued with
119\,930\,299\,362 data points between them.

Spatially, the data from the Northern hemisphere cover the sky north
of +20 deg and south of +66 deg, in the Southern hemisphere it is
south of $-$20 deg almost to the pole. Temporally the data begin in the North on the 2nd
May 2004 and go through to the 9th August 2008, in the South the data
range from 13th February 2006 to 27th May 2008 (see
Table~\ref{summary}). 

\begin{table}
\begin{center}
\caption{Summary of the data.}
\label{summary}
\begin{tabular}{lr}
\hline
\hline
Number of unique objects                      & 17\,970\,937\\
Number of raw science images                  &  3\,631\,972\\
Number of flat field frames                   & 257\,267\\
Number of bias frames                         & 119\,175\\
Number of thermal frames                      & 71\,302\\
Number of data points                         & 119\,930\,299\,362\\
Average number of data points per light curve & 6674\\
Maximum number of data points                 & 74\,013\\
Light curve disk space                        & 5.0 TB\\
Raw science image disk space                  & 14.5 TB\\
Raw calibration image disk space              & 1.6 TB\\
\hline
Start date (North)                            & 2/5/2004\\
End date (North)                              & 9/8/2008\\
Longitude (North)                             & $-$17:52:45\\
Latitude (North)                              & 28:45:37\\
Minimum declination (North)                   & +20\\
Maximum declination (North)                   & +65.86\\
\hline
Start date (South)                            & 13/2/2006\\
End date (South)                              & 27/5/2008\\
Longitude (South)                             & 20:48:38\\
Latitude (South)                              & $-$32:22:51\\
Minimum declination (South)                   & $-$20\\
Maximum declination (South)                   & $-$89.07\\
\hline
\hline
\end{tabular}
\end{center}
\end{table}

\section{Data products}
\label{data_products}

\subsection{Light curves}

\begin{figure}
  \includegraphics[width=\hsize]{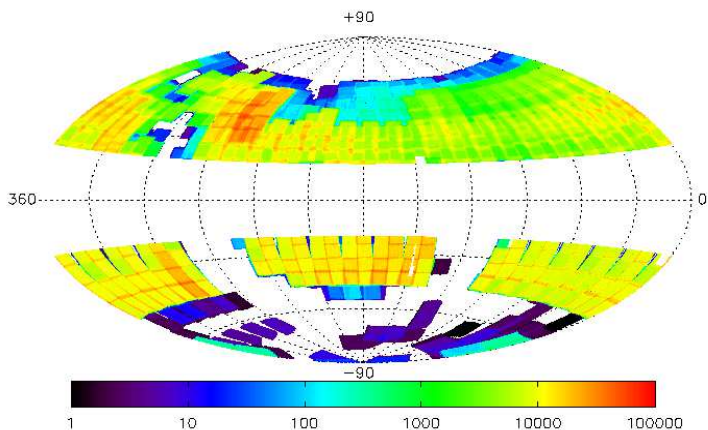}
  \caption{Hammer-Aitoff projection of the average number of data
    points per light curve available in this data release.}
  \label{avgpoints}
\end{figure}

The main data product is that of light curves of individual
objects. Each individual light curve is stored in binary FITS table
format. Some of these have data spanning back to 2004 while others
may have just one seasons data. The average number of data points in
the light curves is $\sim$6600, while some have more than 70\,000 (see
Fig.~\ref{avgpoints} and Table~\ref{summary}).

\begin{figure}
  \includegraphics[width=\hsize]{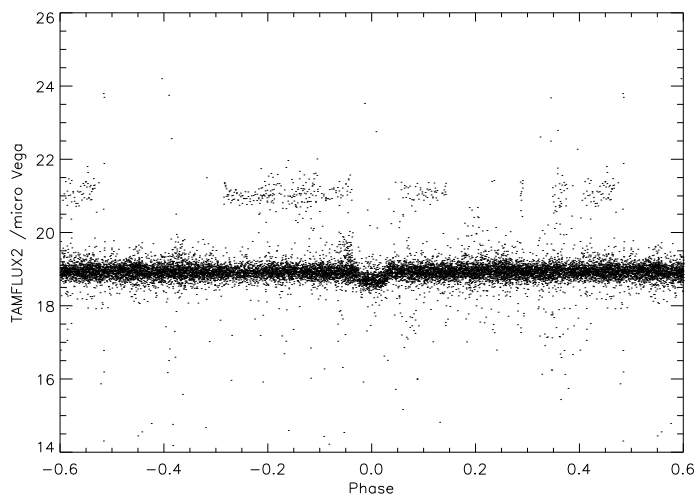}
  \caption{1SWASP J002040.07+315923.7 (WASP 1) data taken from this
    data set and folded at its orbital period (2.51997 days).}
  \label{wasp1}
\end{figure}

Table~\ref{lc_columns} lists the columns in the light curve FITS
file. Each row corresponds to a data point in the light curve and is
the result of aperture photometry of the
images by the pipeline, as described in \citet{kane04}. This
results in a processed flux measurement (FLUX2) of each object from
each image. Further to this flux, a TAMUZ corrected flux is given, which gives a
more consistent measurement between different cameras and
years \citep[][(Cameron refers to the TAMUZ correction as
  SYSREM)]{cameron06}. Fig.~\ref{wasp1} has an example of a TAMUZ
corrected light curve of WASP~1 constructed from the data available
here. 

Also included in the light curve data are the position of the object on the CCD,
the unique image ID that the data point was derived from, and a flag
to indicate if a TAMUZ correction has been made or not. The TAMUZ flag is a bit-mask; if the
correction has been made then this will be 32, if it is zero then the
TAMUZ flux and error will be a copy of the FLUX2 values.

TMID is the heliocentrically corrected mid-point of the exposure in
seconds after 2004-01-01T00:00:00, and can be converted to HJD using

\begin{equation}
HJD = TMID/86\,400 + 2453005.5
\end{equation}

Along with the basic photometry, additional meta-data can be
requested, for each data point this is added to the light curves on
the fly. These include the airmass and general weather conditions (see
Table~\ref{lc_columns}).

\begin{table*}
\begin{center}
\caption{Data available for each photometric point.}
\label{lc_columns}
\begin{tabular}{lll}
\hline
\hline
Name               & Unit                & Description\\
\hline
\multicolumn{3}{c}{Stored in the light curves on disk}\\
\hline
TMID               & sec                 & Mid-time of exposure\\
FLUX2              & micro Vega          & Processed flux\\
FLUX2\_ERR         & micro Vega          & Processed flux error\\
TAMFLUX2           & micro Vega          & TAMUZ corrected processed flux \\
TAMFLUX2\_ERR      & micro Vega          & TAMUZ flux error\\
CCDX               & 1/16th of pixel     & X position on the CCD \\
CCDY               & 1/16th of pixel     & Y position on the CCD \\
IMAGEID            & --                  & Unique image ID \\
FLAG               & --                  & Bitmask \\ 
\hline
\multicolumn{3}{c}{Added to the light curves on the fly (if requested)}\\
\hline
AIRMASS            & --                  & Airmass\\
ZENDIST            & degrees             & Zenith distance\\ 
MOONALT            & degrees             & Degrees above horizon \\
MOONDIST           & degrees             & Degrees from image centre \\
MOONPHAS           & \% of full moon     & Moon phase\\
WXTIME             & YYYY-MM-DD HH:MM:SS & Time the weather data was taken \\
WXTEMP             & degrees C           & Ambient air temperature \\
WXPRES             & milliBars           & Atmospheric pressure \\
WXWNDSPD           & kph                 & Wind speed \\
WXWNDDIR           & degrees E of N      & Wind direction \\
WXHUMID            & \%                  & Outdoor humidity \\
\hline
\hline
\end{tabular}
\end{center}
\end{table*}

In both the flux cases the units are given as micro Vegas, this
gives a simple conversion to magnitude given by

\begin{equation}
mag = 15 - 2.5 \log(flux)
\end{equation}

where $flux$ is in micro Vegas. This implies a flux of 1.0 micro Vega
corresponds to 15th mag, and $10^6$ micro Vegas; 0th mag. An average
magnitude is calculated for each object by cutting the flux at four
sigma from the median then taking the mean.

The nomenclature used for the target names is \texttt{1SWASP
  Jhhmmss.ssSddhhmmss.s}, every source follows this naming convention
and is therefore the unique identifier used.

\subsection{Images}
\subsubsection{Science frames}
The raw science images are also offered as a data product, these are
stored in FITS image format. Each image is 2048$\times$2048 pixels
with a field of view of 7.8$\times$7.8 degrees and has an exposure
time of 30~s. The naming convention of the images is three digits
corresponding to the CCD number followed by the date and time.

WCS meta-data is stamped into the image headers by the astrometry part
of the main reduction pipeline. There are however two main instances
where this is not the case. Early in pipeline development this information was not logged,
so some of the 2004 data are missing this information. The pipeline
assesses the quality of each image before processing it, some images
are therefore rejected by the pipeline at this stage due to e.g. satellite
trails. Since this will not affect the whole image we have chosen to
leave these images in the public archive.

\subsubsection{Calibration frames}
The calibration frames are also available for download to allow the
reduction of the raw images. These are taken at dusk and dawn most
nights. If the weather conditions are bad at dusk and dawn on a given
night, it is possible that science frames will be available, but no
calibration frames will be present.

The calibration frames follow the same naming convention as the
science frames, but can be distinguished by the \texttt{IMGTYPE}
keyword in the header.

\subsection{Version control}
As the archive evolves the software used to extract and serve the data
will likely be developed further. Since some of the data undergoes
on-the-fly processing as it is served, software development may alter
the final data products. To keep track of this the headers of the FITS
files keep a record of what piece of software has edited it and when. A
version log of the software on the web pages can then be used to see
if software upgrades have affected the science data.

\section{The interface}
\label{interface}

All the data is available via a simple web based interface at
\texttt{www.wasp.le.ac.uk/public/}.

\subsection{Light curves}

These can be found by querying a region of the sky with a specified
R.A. and declination, along with a magnitude range and a minimum number of
data points. This returns a list of all the objects close-by, along
with their position, magnitude and the number of data
points. Individual light curves can be downloaded, or multiple light
curves selected and downloaded as a single operation. 

The extra meta-data in Table~\ref{lc_columns} can be added to individual light
curves on-the-fly at this stage, however, this is a complex process
and slows the retrieval process significantly.

Individual light curves have unique URLs which are made up of
concatenating the object name with
\texttt{www.wasp.le.ac.uk/public/}, so e.g. 1SWASP
J002040.07+315923.7 can be retrieved via
\texttt{www.wasp.le.ac.uk/public/1SWASP J002040.07+315923.7.fits}.

\subsection{Science and calibration images}
The science images can be queried in a similar way to the light curves and a
region of the sky searched. Since the image locations are indexed on
the centre of the image, a search radius of 5.52 degrees (the diagonal
length of the image) is used. The science images can also be queried with a
light curve object ID, this allows all of the images that were used to
generate a light curve to be found and downloaded.

In an analogous way to the light curves, the raw images have unique
URLs, e.g. image 103200405270502090 can be retrieved via
\texttt{www.wasp.le.ac.uk/public/103200405270502090.fits}.

The availability of the calibration images is summarised in table
format, which allows nightly or monthly downloads on a per camera
basis. 

\subsection{Virtual observatory}
An index of the the light curve data is available in the virtual
observatory with the title `1SWASP\_LIGHTCURVES: SuperWASP lightcurves
DR1' published by LEDAS\footnote{\texttt{www.ledas.ac.uk}}. This
allows cross-correlation with other catalogues and the
easy retrieval of data products which can then be analysed with the
standard virtual observatory tools. 

\section{Conclusion}

This first data release (DR1) of the WASP archive has over 3.5 million
images and almost 18 million light curves covering a large
fraction of the sky. All of this is readily available for download
from the web interface (\texttt{www.wasp.le.ac.uk/public/}) or in the
virtual observatory.

Over the coming years, as the data set grows, more data will be added to
the public archive covering a larger part of the sky and with a longer
temporal baseline.

\begin{acknowledgements}
The WASP Consortium consists of astronomers primarily
from the Queen’s University Belfast, Keele, Leicester, The Open University,
St Andrews, the Isaac Newton Group (La Palma), the Instituto de Astrofisica
de Canarias (Tenerife) and the South African Astronomical Observatory. The
WASP-N and WASP-S Cameras were constructed and are operated with
funds made available from Consortium Universities and the UK‘s Science and
Technology Facilities Council.
\end{acknowledgements}

\bibliographystyle{aa}
\bibliography{15655ref}

\begin{thebibliography}{7}
\expandafter\ifx\csname natexlab\endcsname\relax\def\natexlab#1{#1}\fi

\bibitem[{{Collier Cameron} {et~al.}(2009){Collier Cameron}, {Davidson},
  {Hebb}, {Skinner}, {Anderson}, {Christian}, {Clarkson}, {Enoch}, {Irwin},
  {Joshi}, {Haswell}, {Hellier}, {Horne}, {Kane}, {Lister}, {Maxted}, {Norton},
  {Parley}, {Pollacco}, {Ryans}, {Scholz}, {Skillen}, {Smalley}, {Street},
  {West}, {Wilson}, \& {Wheatley}}]{cameron09}
{Collier Cameron}, A., {Davidson}, V.~A., {Hebb}, L., {et~al.} 2009, MNRAS,
  400, 451

\bibitem[{{Collier Cameron} {et~al.}(2006){Collier Cameron}, {Pollacco},
  {Street}, {Lister}, {West}, {Wilson}, {Pont}, {Christian}, {Clarkson},
  {Enoch}, {Evans}, {Fitzsimmons}, {Haswell}, {Hellier}, {Hodgkin}, {Horne},
  {Irwin}, {Kane}, {Keenan}, {Norton}, {Parley}, {Osborne}, {Ryans}, {Skillen},
  \& {Wheatley}}]{cameron06}
{Collier Cameron}, A., {Pollacco}, D., {Street}, R.~A., {et~al.} 2006, MNRAS,
  373, 799

\bibitem[{{Kane} {et~al.}(2004){Kane}, {Collier Cameron}, {Horne}, {James},
  {Lister}, {Pollacco}, {Street}, \& {Tsapras}}]{kane04}
{Kane}, S.~R., {Collier Cameron}, A., {Horne}, K., {et~al.} 2004, MNRAS, 353,
  689

\bibitem[{{Monet} {et~al.}(2003){Monet}, {Levine}, {Canzian}, {Ables}, {Bird},
  {Dahn}, {Guetter}, {Harris}, {Henden}, {Leggett}, {Levison}, {Luginbuhl},
  {Martini}, {Monet}, {Munn}, {Pier}, {Rhodes}, {Riepe}, {Sell}, {Stone},
  {Vrba}, {Walker}, {Westerhout}, {Brucato}, {Reid}, {Schoening}, {Hartley},
  {Read}, \& {Tritton}}]{monet03}
{Monet}, D.~G., {Levine}, S.~E., {Canzian}, B., {et~al.} 2003, AJ, 125, 984

\bibitem[{{Norton} {et~al.}(2007){Norton}, {Wheatley}, {West}, {Haswell},
  {Street}, {Collier Cameron}, {Christian}, {Clarkson}, {Enoch}, {Gallaway},
  {Hellier}, {Horne}, {Irwin}, {Kane}, {Lister}, {Nicholas}, {Parley},
  {Pollacco}, {Ryans}, {Skillen}, \& {Wilson}}]{norton07}
{Norton}, A.~J., {Wheatley}, P.~J., {West}, R.~G., {et~al.} 2007, A\&A, 467,
  785

\bibitem[{{Pollacco} {et~al.}(2006){Pollacco}, {Skillen}, {Cameron},
  {Christian}, {Hellier}, {Irwin}, {Lister}, {Street}, {West}, {Anderson},
  {Clarkson}, {Deeg}, {Enoch}, {Evans}, {Fitzsimmons}, {Haswell}, {Hodgkin},
  {Horne}, {Kane}, {Keenan}, {Maxted}, {Norton}, {Osborne}, {Parley}, {Ryans},
  {Smalley}, {Wheatley}, \& {Wilson}}]{pollacco06}
{Pollacco}, D.~L., {Skillen}, I., {Cameron}, A.~C., {et~al.} 2006, PASP, 118,
  1407

\bibitem[{{Wilson} {et~al.}(2008){Wilson}, {Gillon}, {Hellier}, {Maxted},
  {Pepe}, {Queloz}, {Anderson}, {Collier Cameron}, {Smalley}, {Lister},
  {Bentley}, {Blecha}, {Christian}, {Enoch}, {Haswell}, {Hebb}, {Horne},
  {Irwin}, {Joshi}, {Kane}, {Marmier}, {Mayor}, {Parley}, {Pont}, {Ryans},
  {Segransan}, {Skillen}, {Street}, {Udry}, {West}, \& {Wheatley}}]{wilson08}
{Wilson}, D.~M., {Gillon}, M., {Hellier}, C., {et~al.} 2008, ApJL, 675, L113

\end{thebibliography}

\end{document}